\documentclass[11pt]{article}
\usepackage[a4paper, total={6in, 8in}]{geometry}
\usepackage{amsmath}
\usepackage{amsthm}
\usepackage{amsfonts}
\usepackage{bm}
\usepackage{graphicx}
\usepackage{natbib}
\RequirePackage[colorlinks,linkcolor=blue,citecolor=blue,urlcolor=blue]{hyperref}
\usepackage{siunitx}
\usepackage{booktabs}
\usepackage{subcaption}
\usepackage[mathscr]{euscript}
\usepackage{algorithm,algpseudocode}
\usepackage{setspace}
\usepackage{breqn}
\usepackage{multirow}
\usepackage{enumitem}
\usepackage{hyperref}
\usepackage{enumitem}
\counterwithin{figure}{section}
\counterwithin{table}{section}

\newcommand{\Tset}{\bm{\mathscr{T}}} 
\newcommand{\Cset}{\bm{\mathscr{C}}} 

\title{Protocol for an Observational Study on the Effects of Giving Births from Unintended Pregnancies on Later Life Physical and Mental Health}
\author{
  Samrat Roy\\
  \small{Department of Statistics and Data Science, University of Pennsylvania}\\
   \small\textit{{email: \href{mailto:roysa@wharton.upenn.edu}{roysa@wharton.upenn.edu}}}
  \and
  Marina Bogomolov\\
  \small{Industrial Engineering and Management, Technion - Israel Institute of Technology}\\
   \small\textit{{email: \href{mailto:marinabo@technion.ac.il}{ marinabo@technion.ac.il}}}
  \and
  Ruth Heller\\
  \small{Department of Statistics and Operations Research, Tel-Aviv University}\\
   \small\textit{{email: \href{mailto:ruheller@tauex.tau.ac.il}{ ruheller@tauex.tau.ac.il}}}
   \and
  Amy M. Claridge\\
  \small{ Child Development and Family Science, Central Washington University}\\
  \small{\textit{email: \href{mailto:claridgea@cwu.edu}{claridgea@cwu.edu}}}
  \and
  Tishra Beeson\\
  \small{ Department of Health Sciences, Central Washington University}\\
  \small{\textit{email: \href{mailto:Tishra.Beeson@cwu.edu}{Tishra.Beeson@cwu.edu}}}
  \and
  Dylan S. Small\\
  \small{Department of Statistics and Data Science, University of Pennsylvania}\\
  \small{\textit{email: \href{mailto:dsmall@wharton.upenn.edu}{dsmall@wharton.upenn.edu}}}
}
\date{June 2022}

\begin{document}

\maketitle
\begin{abstract}
    There has been increasing interest in studying the effect of giving births to unintended pregnancies on later life physical and mental health. In this article, we provide the protocol for our planned observational study on the long-term mental and physical health consequences for mothers who bear children resulting from unintended pregnancies. We aim to use the data from the Wisconsin Longitudinal Study (WLS) and examine the effect of births from unintended pregnancies on a broad range of outcomes, including mental depression, psychological well-being, physical health, alcohol usage, and economic well-being. To strengthen our causal findings, we plan to address our research questions on two subgroups, Catholics and non-Catholics, and discover the “replicable” outcomes for which the effect of unintended pregnancy is negative (or, positive) in both subgroups. Following the idea of non-random cross-screening, the data will be split according to whether the woman is Catholic or not, and then one part of the data will be used to select the hypotheses and design the corresponding tests for the second part of the data.  In past use of cross screening (automatic cross screening) there was only one team of investigators that dealt with both parts of the data so that the investigators would need to decide on an analysis plan before looking at the data.  In this protocol, we describe our analysis plan for carrying out automatic cross screening to study the effects of unintended pregnancy.  In addition, we describe plans to carry out a novel flexible cross-screening in which there will be two teams of investigators with access only to one part of data and each team will use their part of the data to decide how to plan the analysis for the second team's data. In addition to the above replicability analysis, we also discuss the plan to test the global null hypotheses, in order to identify  outcomes which are affected by unintended pregnancy for at least one of the two subgroups of Catholics and non-Catholics.  
\end{abstract}

\section{Background and Motivation}
Unintended pregnancies are prevalent in the United States (U.S.) and worldwide. A pregnancy may be considered unintended when the pregnancy was not wanted, or when it was mistimed or earlier than desired (CDC, 2021). $45\%$ of all U.S. pregnancies were unintended in 2011 (\cite{finer2016declines}), with higher rates of unintended pregnancies among adolescents and young adults (\cite{postlethwaite2010pregnancy}), among women living at or below poverty (\cite{finer2016declines}), among unmarried individuals (\cite{musick2002planned}) and among non-Hispanic black and African American women (\cite{finer2016declines}). Unintended pregnancy rates have been estimated at $44\%$, worldwide (\cite{bearak2018global}). 

There has been an escalated interest in studying the effect of unintended pregnancies on the later life mental and physical health of the women (\cite{bahk2015impact}, \cite{herd2016implications}, \cite{barton2017unplanned}). This effect can either be attributed to the termination of the unintended pregnancy, or to the continuation of the unintended pregnancy to term. While the former has been discussed by a robust body of literature (see \cite{herd2016implications} and references therein), there has not been much developments on the latter one. Birthing individuals with unintended pregnancies tend to report less emotional attachment to their baby in pregnancy (\cite{pakseresht2018physical}), needing more time to accept their pregnancy. After birth, parents of unintended pregnancies are less likely to initiate breastfeeding (\cite{mark2022pregnancy}), tend to report less expression of affection (\cite{hayatbakhsh2011longitudinal}) and issues in attachment formation (\cite{miller2009preconception}). Also, they are more likely to report parenting stress at 6 months, one year postpartum, and to engage in less effective parenting strategies (\cite{miller2009preconception}, \cite{east2012adolescents}). Thus, a robust large-scale study on the physical and mental health consequences of the mothers who continued their unintended pregnancies to term is of critical importance. Most of the existing literature in this regard, considered the births only after \textit{Roe v. Wade} ($1973$)\footnote{see \url{https://en.wikipedia.org/wiki/Roe_v._Wade}}, which protected the liberty of a pregnant woman to choose to have an abortion, and $40\%$ of the post-\textit{Roe v. Wade} ($1973$) unintended pregnancies were terminated. Thus, the existing literature, that consider the births only after \textit{Roe v. Wade} ($1973$), can hardly capture the actual effect of unintended pregnancies on later life physical and mental health. This gap in the literature, along with the recent overturning of \textit{Roe v. Wade} by the Supreme Court\footnote{see \url{https://www.npr.org/2022/06/24/1102305878/supreme-court-abortion-roe-v-wade-decision-overturn}}, raises the question, ``What are the long-term mental and physical health
consequences for mothers who bear children
resulting from unintended pregnancies?".

In this study, we address the above question by using data from Wisconsin Longitudinal Study (WLS) (\cite{herd2014cohort}), wherein the respondents were the women who graduated from Wisconsisn High School in $1957$, and the survey data on various aspects of those respondents' life course were collected in $1957, 1964,1975, 1992, 2004$ and $2011$. \cite{herd2016implications} used the same data and analyzed the later life physical and mental health consequences of the women who gave births to unintended pregnancies. As they pointed out, the WLS data has some advantageous features which alleviate the aforementioned drawbacks in the existing literature. First, unlike the previous work on the effects of unwanted pregnancies, WLS respondents had experienced nearly all their pregnancies before the 1973 \textit{Roe v. Wade} decision. Thus, most, if not all, of these women did not have the opportunity to
terminate an unintended pregnancy and hence this data is more reliable for inferring  the actual effect of giving births to unintended pregnancies on later life physical and mental health. Second, WLS data consists of a wide range of covariates, including the family
background, adolescent characteristics, educational and occupational achievement and aspirations, which
could potentially confound the relationship between
unplanned pregnancies and later-life mental and physical health outcomes. We use these variables to create matched set of treatment and control individuals and then compare the physical and mental health outcomes within each matched set. This is a standard practice as performing a randomized control study is highly unfeasible in this case, and we need to depend solely on the observational data. Finally, this data, unlike the other relevant ones, tracks the information longitudinally at multiple time points, and this  facilitates the study of later life physical and mental health consequences. 

We suggest our own statistical design in order to answer the research questions. Our suggested design addresses potential biases by planning to carry out both a replicability analysis and a sensitivity analysis. The replicability analysis will be possible since we address the research questions on two subgroups, Catholics and non-Catholics. More specifically, we aim to discover the outcomes for which the effect of unintended pregnancy is negative in both Catholic and non-Catholic subgroups,
as well as the outcomes for which the effect is positive in both subgroups. These outcomes are usually referred to as ``\textit{replicable}" findings (\cite{Bogomolov2022ReplicabilityAM}). When treatments are not randomly assigned, the evidence that the treatment is the cause of its ostensible effects is strengthened by showing that people who receive the treatment for different reasons experience similar effects (\cite{rosenbaum2015see}). Catholics and non-Catholics may have had unintended pregnancies for somewhat different reasons. The Catholic Church opposes birth control while most other faiths do not. Consequently, Catholic women who have unintended pregnancies are relatively more likely to have them because their religious beliefs forbid birth control while non-Catholic women are relatively more likely to have unintended pregnancies because they chose not to use birth control. We employ the idea of \textit{directional replicability}, introduced in \cite{repl_biometrika}, in order to achieve this goal (see Section \ref{testing} for more details). We use \textit{non-random cross-screening}, wherein, the data is split according to whether the woman is Catholic or not. One part of the data is used to select the hypotheses to be tested based on the second part of the data, and then to design the corresponding tests. The scenario, wherein, the same team of investigators deals with both parts of the data, is referred to as ``\textit{Automated cross-screening}". In such cases, to guarantee family wise error rate (FWER) control, the investigators need to decide before seeing the data how they will use one part of the data to design the analysis of the second part. On the contrary, when there are two teams of investigators, each having access only to one part of the data, then each team can see their part of the data and decide how to plan the analysis for the second team's data. This is referred to as ``\textit{Flexible cross-screening}". In Section \ref{testing}, we provide a detailed description on how we use these non-random cross-screening methods to identify the replicable outcomes with positive (or, negative) treatment effect in both the Catholic and the non-Catholic subgroups. In addition to the above replicability analysis, we also discuss the plan to test the global null hypothesis in Appendix \ref{global}, that is intended to identify the outcomes which are affected by unintended pregnancy for at least one of the two subpopulations of Catholics and non-Catholics. \\
\\
The remainder of the document is organized as follows. Section \ref{matching} discusses the matching method that we use to prepare the matched set of treated and control individuals for both the Catholics and non-Catholics subgroups. In Section \ref{outcomes}, we summarize the later-life mental, physical and economic outcomes on which we test the effect of unintended pregnancies. Section \ref{testing} and Appendix \ref{global} discuss the details of our proposed testing design for replicability analysis and global null respectively. Finally we conclude with some simulation studies in Appendix \ref{simulation}, which is followed by some additional tables in Appendix \ref{add_tebs}.

\section{Risk-set matching}\label{matching}
We employ \textit{Risk Set Matching} (\cite{li2001balanced,zubizarreta2014isolation}) separately for both the Catholic and non-Catholic women, and deal with the potential confounding in each subgroup. Note that different women gave births to unintended pregnancies at different years. Thus, unlike the randomized experiments, where all the subjects are assigned to treatment or control at some fixed time point, in this case, there is no such fixed time of treatment assignment. While some women can give birth to unintended pregnancies now (and are assigned to the treatment group now), others can give births to unintended pregnancies years later (and will be assigned to the treatment group then) or, may not have it at all. Also, the individuals not giving birth to any unintended pregnancies yet (thus currently assigned to the control group), may end up giving births to unintended pregnancies later and then get assigned to the treatment group. Risk-set matching, proposed by \cite{li2001balanced}, enables us to deal with the aforementioned temporal structure of the treatment assignment. The matching algorithm acts sequentially over time, and at each time point it pairs two individuals from two different groups. The first group of subjects are the ones who just got treated for the first time at that time point, while the second group consists of the ones who have not been treated yet. While pairing the individuals, the matching is performed based on the observed covariates prior to that time point, ensuring that the two individuals looked similar in terms of covariates till the point when one of them received treatment for the first time, and the other one did not receive the treatment yet. The term `risk-set' relates to the Cox's proportional hazards model, where the two individuals will have `similar' time-dependent covariates and thus will be at similar `risk' of receiving the treatment just before the moment one of them actually receives it.

What follow next, are the description of the covariates based on which the women were matched and then the detailed steps of the risk-set matching.\\ 
\\
\textbf{\textit{Covariates}}: The women are matched on some time-varying covariates and some fixed covariates as listed below.\\
\\
\noindent \textbf{\textit{Childhood measures}}: We consider the following four childhood measures as fixed covariates for matching. 1) \textit{High School percentile rank}: woman's high school percentile rank on the basis of high school grades, derived as, $100 - [\text{rank in
class} / (\text{no. of students in class} \times 100)]$\footnote{see \url{https://www.ssc.wisc.edu/wlsresearch/documentation/waves/?wave=wls5764&module=aedu}}. 2) \textit{IQ}: derived from an administrative measure, namely, the Henmon-Nelson Test of Mental Ability\footnote{see \url{https://www.ssc.wisc.edu/wlsresearch/documentation/appendices/G/cor652.asc}}, that was administered to all high school students in Wisconsin. The year women took the test varied over time, but most scores employed in the WLS are from the student’s junior year or are adjusted to reflect what their junior scores would be. These scores were then renormed to IQ equivalents on the basis
of the percentile distribution of scores that
were observed among all Wisconsin high
school juniors. 3) \textit{Parental
socioeconomic status}: factor-weighted SES scores for parents \footnote{see \url{https://www.ssc.wisc.edu/wlsresearch/documentation/waves/?wave=wls5764&module=tax}} that are derived based on the following variables: father's education, mother's education, father's
occupation, and the 4-year average of parental income derived from Wisconsin tax records \footnote{see \url{https://www.ssc.wisc.edu/wlsresearch/documentation/appendices/L/cor689.asc}}. 4) \textit{Population of town}: $1957$ population of the town in which the woman attended her high school. Matching on these variables ensures that two women had similar childhood measures before one of them gave birth to an unintended pregnancy for the first time and the other one did not.\\
\\
\noindent \textbf{\textit{Adulthood measures}}: Here we consider the following five time-varying covariates for matching: 1) \textit{Number of children}: this is a time-varying covariate which is used to match the women based on the number of children they had prior to the time point when one of them gave birth to an unintended pregnancy for the first time and the other one did not give any such births yet. For example, if a woman had an unintended pregnancy for her third child, whereas the previous two children were planned, then we would be seeking to match her to another woman who has had two planned pregnancies. 2) \textit{Age}: this is also a time-varying covariate, which is used to match a woman with her first ever birth to an unintended pregnancy to another woman of similar age at that time and who did not have any unintended pregnancies yet. 3) \textit{Marital status}: this time-varying binary covariate that takes the value $1$, if the woman was married at that time point. Thus this can be used to match a woman with her first birth to an unintended pregnancy to another woman of same marital status at that time and with no births to unintended pregnancies till then. 4) \textit{Years of education}: this time-varying covariate captures the years of educational attainment of the women across time. As defined in \cite{zubizarreta2014isolation}, we derive the years of education as $\text{min} (E, A-6)$, where $A$ is the age of the woman at that time point and $E$ is the equivalent years of regular education\footnote{see \url{https://www.ssc.wisc.edu/wlsresearch/documentation/waves/?wave=wls75&module=cedu}} from WLS data library, collected in $1975$. The term $A-6$ relates to the fact that, in the US, the students are usually admitted to the school at the age of $6$ years and they complete high school with $12$ years of education at age $18$. Thus, if a woman still continues  her education at the time of interest, then her total number of years of education will be taken as $A-6$. On the other hand, if the woman had completed her education earlier than the time of interest, then her total number of years of education would be taken as $E$. In our case, $A$ is the woman's age at the time point when one of the two women who are being matched, had her first birth to an unintended pregnancy and the other one did not have any such births yet. Thus matching on $\text{min}(E, A-6)$ ensures that the two women had similar years of education before they were paired. 5) \textit{Whether the first depression occurred on or before the start of pregnancy}: this is a binary time-varying covariate which aims to match the women based on whether she had her first depression before the start of the pregnancy. Since it is hard to find exact data on start of pregnancy and time of first depression, we construct this as a binary variable that takes the value $1$, if the woman had her first depression on or before the time point $(t-2 \text{ years})$, where $t$ is the year of matching wherein one of the women had her first birth to an unintended pregnancy and the other woman did not have any such births yet. \\
\\
\noindent \textbf{\textit{Personality measures}}: These are the five widely used scales of personality, namely, agreeableness, conscientiousness, extraversion, neuroticism, and openness (\cite{john1999big}), which we use as fixed covariates for matching. These measures were first
collected in a 1992 mail survey\footnote{see \url{https://www.ssc.wisc.edu/wlsresearch/documentation/waves/?wave=wls92&module=mpers}}. However, there are studies (\cite{herd2016implications}) which argued that the personality remains fairly stable in pregnancy and the transition to parenthood. Thus even though these measures were recorded post-treatment in 1992, these could be effectively used as pre-treatment covariates for pairing the women based on personality.\\ 
\\
\textbf{\textit{Risk-set matching}}: 
Let $c_i$ be the total number of children for the $i^{th}$ woman. If $c_i$ is greater than zero, then we use the notation $\{y_{i1}, y_{i2},\cdots,y_{ic_i}\}$ to denote the birth years of the $c_i$ children of the $i^{th}$ woman. Similarly, the pregnancy intention for the $c_i$ births of that woman is denoted by $\{v_{i1}, v_{i2},\cdots,v_{ic_i}\}$, where $v_{ij}$ takes the value $1$ for unintended pregnancy, and $0$ otherwise. As mentioned earlier, risk-set matching is performed sequentially over the birth years. For a specific year, it first identifies the women who had given births to unintended pregnancies for the first time in that year. Then it pairs each of those women to another woman with similar covariates and who did not give births to any unintended pregnancies yet. Once the matching is done for a particular year, the matched pairs are removed before moving on to the next year. More specifically, we repeat the following three steps iteratively to obtain the matched pairs:\\
\\
\noindent \textit{Step-1}: Given an iteration, this step finds the earliest year in which at least one birth has occurred from an unintended pregnancy. Let $n$ denote the number of women present at that iteration. Define 
a set $I$ of paired indices as, $I=\{(i,j) \in \{1,2,\cdots,n\} \times \{1,2,\cdots,c_i\} |\text{ }c_i>0  \ \textrm{and} \sum_{j=1}^{c_i}v_{ij}>0\}$  and find $t^*= \underset{(i,j)\in I}{\min} \text{ }y_{ij}$. Thus, at any iteration, $t^*$ is the earliest year in which at least one birth has occurred from an unintended pregnancy.\\
\\
\noindent\textit{Step-2}: Given $t^*$ from the first step, this step formally defines the two sets: set of individuals who had given births to unintended pregnancies in the year $t^*$, and the set of individuals who either had no births, or births only from planned pregnancies till the year $t^*$. More specifically, we define, two sets of indices, $\Tset$ and $\Cset$ as follows:
    $\Tset=\{i\text{ }|\text{ }\exists\text{ }j\in\{1,2,\cdots,c_i\}\text{ with } y_{ij}=t^* \text{ and }v_{ij}=1\}$. Thus, $\Tset$ comprises those women who had given births to unintended pregnancies in the year $t^*$. Note that, at any iteration, $t^*$ is the earliest year when any birth occurs due to unintended pregnancies. Thus, the women in the set $\Tset$, will not have any unintended pregnancies prior to the year $t^*$. Otherwise, if they would have unintended pregnancies before year $t^*$, then they would have already been matched and removed before this iteration (see Step-4). Define the other set $\Cset$ as $\Cset=\Cset_1 \cup \Cset_2$, where $\Cset_1=\{i\text{ }|\text{ } v_{ij}\neq1 \text{ for all } j\in\{1,2,\cdots,c_i\} \text{ with }y_{ij}\leq t^*\}$ and $\Cset_2=\{i\text{ }|\text{ }c_{it^*}=0\}$, with $c_{it^*}$ denoting the number of children for the $i^{th}$ woman till the year $t^*$. Thus $\Cset$ consists of those women who either had no births (that is, $\Cset_2$), or births only from planned pregnancies (that is, they belong to $\Cset_1$) till the year $t^*$. Hence, the women in $\Tset$ and $\Cset$ are `similar' in terms of observed histories of unintended pregnancy prior to year $t^*$. The only difference between $\Tset$ and $\Cset$ occurs at the year $t^*$, when the former group gives births to unintended pregnancies, and the latter does not.\\
    \\
    \noindent \textit{Step-3}: In this step, each woman in the set $\Tset$ is individually matched to a woman in $\Cset$ based on the aforementioned covariates. We use robust Mahalanobis distance (robust MD) (\cite{yeager2019using}, \cite{li2019outlier}) for this purpose, which replaces the values of the covariates with their ranks (with average rank for ties), pre-multiplies and post-multiplies the covariance matrix of the ranks by a diagonal matrix whose diagonal elements are the ratios of the standard deviations of untied ranks to the standard deviations of the tied ranks of the covariates and finally computes the Mahalanobis distance using the ranks and the above adjusted covariance matrix. Unlike Mahalanobis distance, robust MD does not suffer from ``outlier masking” (\cite{filzmoser2005multivariate}), where the outliers will have undue influence on the mean and covariance estimation. In addition to the above distance metric, we also add a time-dependent propensity score caliper (\cite{lu2005propensity}) to it as follows. First we fit a time-dependent Cox Proportional Hazards (Cox PH) model (\cite{dr1972regression}, \cite{lu2005propensity}):
\begin{equation}
\label{coxph}
    h_m(t)= h_0(t)\text{ }exp\{\beta^T X_m(t)\}
\end{equation}
where, $h_m(t)$ is the hazard of giving birth to an unintended pregnancy for the $m^{th}$ woman at $t^{th}$ time point. $h_0(t)$ is the baseline hazard and $X_m(t)$ is the vector of all the fixed and time-varying covariates at time point $t$. The time-dependent propensity scores for the $l^{th}$ and the $k^{th}$ individuals are then computed as the fitted hazard from the Cox PH model \eqref{coxph}, denoted by $\hat{\beta}^TX_l(t)$ and $\hat{\beta}^TX_k(t)$ respectively, and then the penalty function $1000 \times \text{max}(0,\lvert \hat{\beta}^TX_l(t) -\hat{\beta}^TX_k(t)\lvert-w)$ is added to the robust MD between that pair. Here the caliper $w$ is chosen as $50\%$ of the pooled within group standard deviation of the propensity scores (\cite{hansen2011propensity}). Thus this function adds penalization to the robust MD when the propensity scores are further apart than the caliper $w$. \\
    \\
    \noindent \textit{Step-4}: Following the previous step, each of the women in set $\Tset$ is matched to another woman from set $\Cset$. Suppose $\{(T_1,C_1), (T_2,C_2), \cdots, (T_s,C_s)\}$ are the $s$ matched pairs where $s$ is the number of women in $\Tset$. Finally, these matched pairs are removed before repeating \textit{Step-1}. We repeat these steps iteratively until there is no woman left who ever had any birth to an unintended pregnancy.\\
    
\begin{center}
\begin{figure}[htbp]
\includegraphics[scale=0.9]{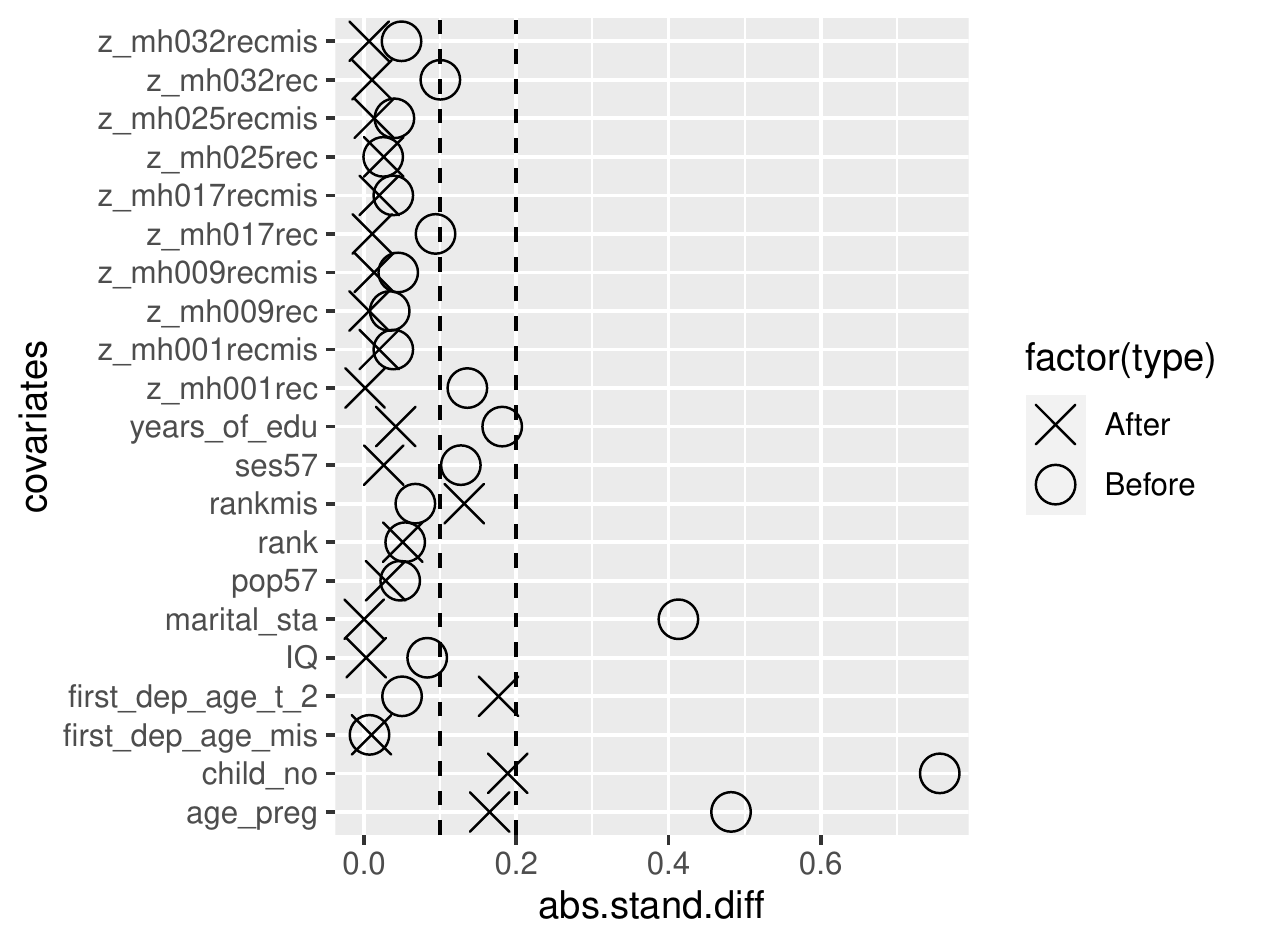}
\caption{Covariate balancing after matching for the Catholics subgroup: all the absolute standardized differences after matching are less than or equal to 0.2.}
\label{lpc}  
\end{figure}  
\end{center}

\begin{center}
\begin{figure}[htbp]
\includegraphics[scale=0.9]{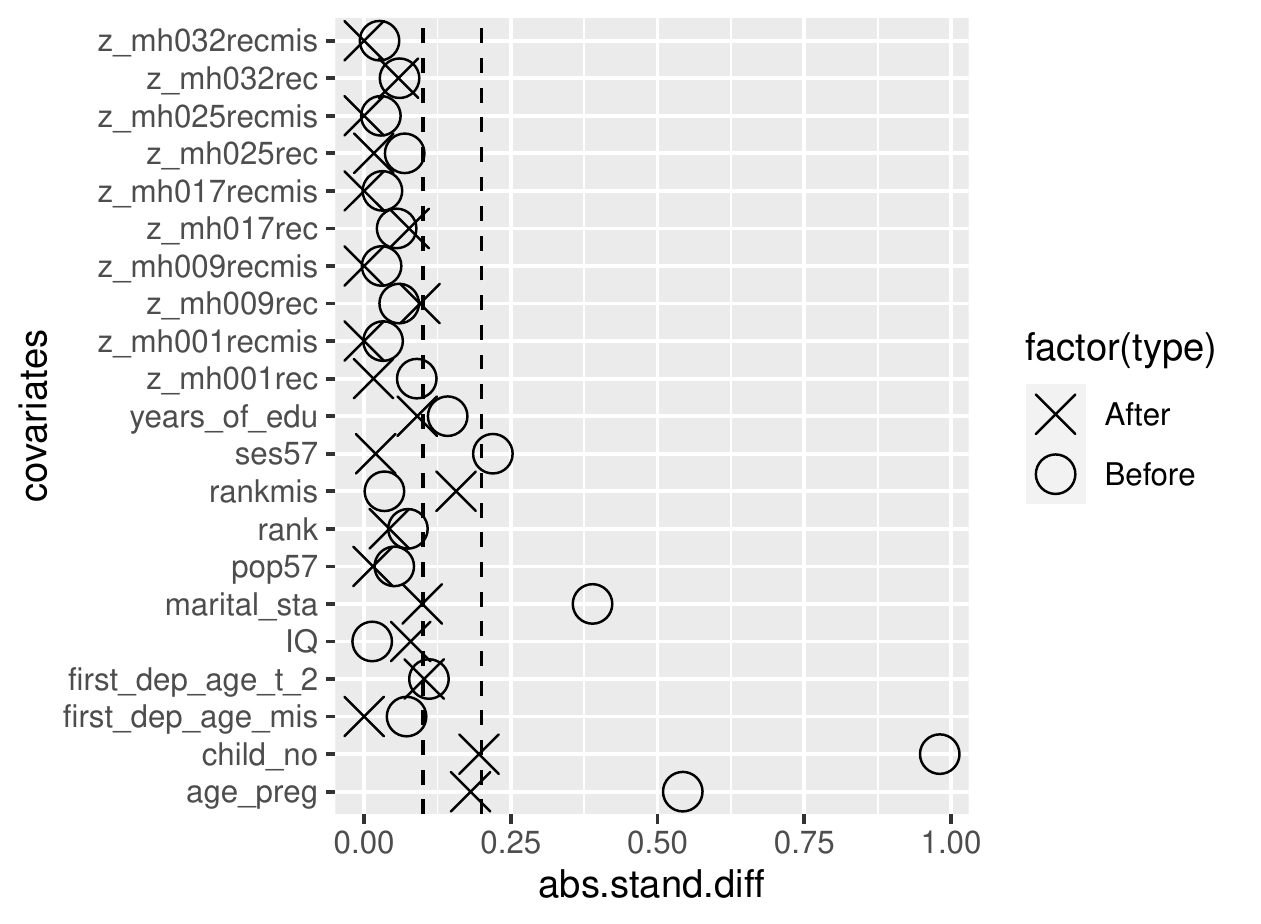}
\caption{Covariate balancing after matching for the Non-Catholics subgroup: all the absolute standardized differences after matching are less than or equal to 0.2.}
\label{lpnc}  
\end{figure}  
\end{center}

    To assess whether the matching has constructed a treated and control group that are balanced on the observed covariates, we use the standardized differences, which are the differences in the means of the covariates between the treated and control group in standard deviation units, and are formally defined as $\frac{\bar{x}_{trt}-\bar{x}_{cont}}{\sqrt{\frac{s^2_{trt}+s^2_{cont}}{2}}}$. Figure \ref{lpc} and Figure \ref{lpnc} depict the absolute values of the pre-matching and post-matching standardized differences for the Catholics and Non-Catholics subgroups respectively. As can be seen, for both the subgroups, the absolute values of the post-matching standardized differences are quite small (less than or equal to 0.2), suggesting a well balance of covariates between the treatment and control group.     

    \section{Later-Life Mental, Physical and Economic Outcomes}
    \label{outcomes}
\begin{figure}
\centering
\includegraphics[scale=0.6,trim=1cm 0.1cm 2cm 4cm]{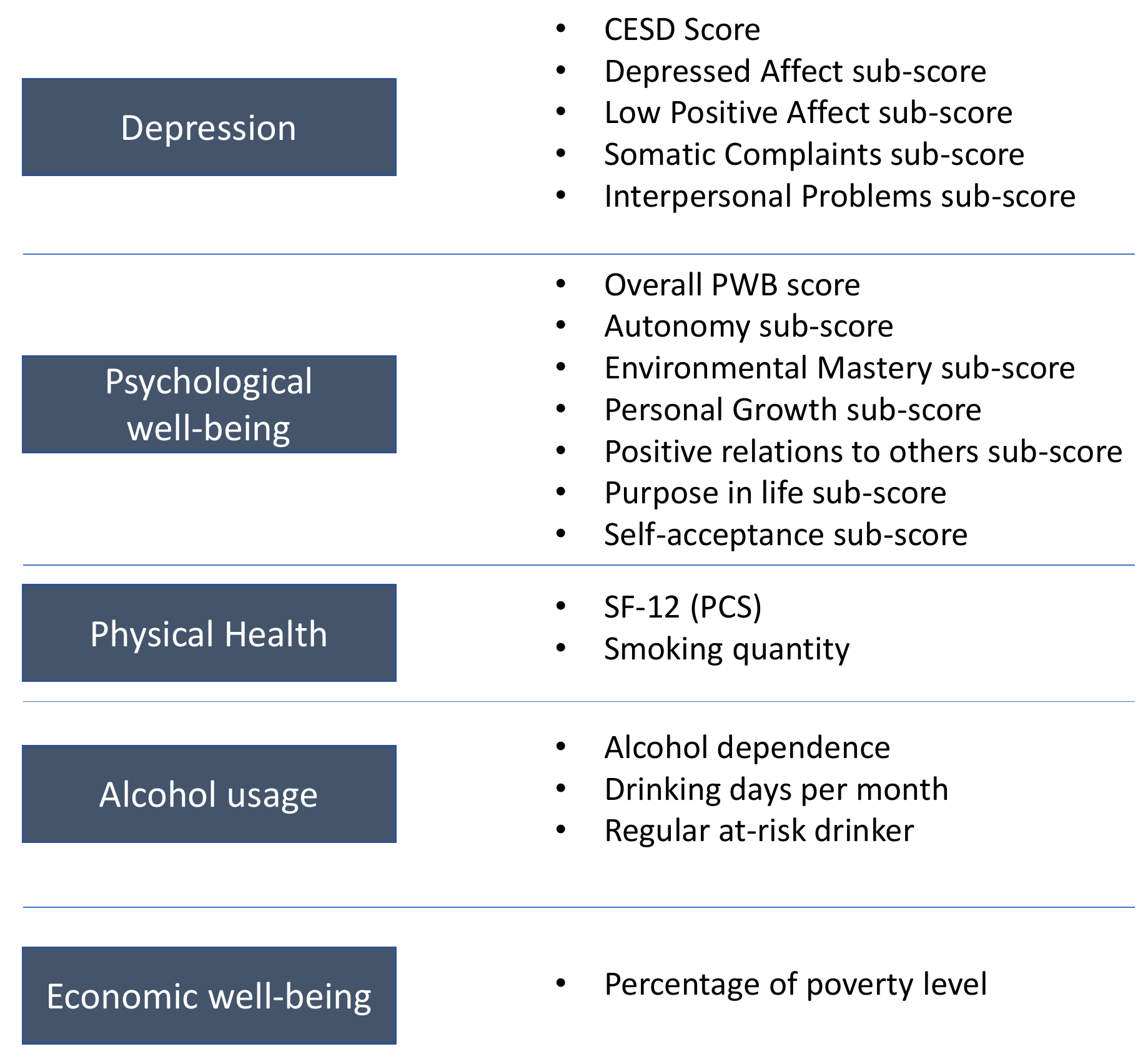}
\caption{List of outcomes for each of the five aspects. }\label{fig:outcomes}
\end{figure}

    Our study primarily focuses on the five aspects of later-life, namely, depression, psychological well-being, physical health, alcohol usage, and economic well-being. For each of these aspects, we consider several outcomes summarized in Figure \ref{fig:outcomes}. For each of the matched pairs obtained in Section \ref{matching}, these outcomes are observed for the survey year 1992-93. What follows next, is a detailed description of the outcomes under each of the five aspects.\\
    \\
    \noindent \textbf{\textit{Depression}}: The outcome that is used to measure the later-life depression is the \textit{Center for Epidemiological Studies-Depression} score (CES-D score). As a depression screening tool, the CES-D contains 20 questions referring to the most common depression symptoms, as defined by the American Psychiatric Association Diagnostic and Statistical Manual (DSM-V). Before administering the questionnaire, the subjects are instructed to think of how frequently (number of days) they have experienced each of the 20 items in the last week and answer accordingly. The summary CES-D score is then calculated as the sum of all those 20 frequencies. Before summing over the frequencies and arriving at the summary score, the frequencies for 4 questions (see questions in the Appendix) need to be recalculated as \textit{correct frequency = 7 - reported frequency}. Unlike the other questions, the wording of these 4 questions relates to opposite symptoms, such as happiness, hopefulness and confidence, and thus the aforementioned adjustment is needed. It is worth mentioning that the above calculation of CES-D in WLS is slightly modified compared to the standard CES-D, which collapses the number of days into four categories, $<1$, 1-2, 3-4, 5-7 and assigns scores $0, 1, 2,$ and $3$ respectively. Thus, while the standard CES-D ranges from $0$ to $60$ (score 3 for each question), the modified one in WLS ranges over $0$ to $140$ (frequency $7$ for each question). In both cases, a higher score implies a higher level of depression.       
    
    In addition to the modified CES-D score, we consider four additional depression outcomes discussed next. \cite{radloff1977ces} performed a principal component factor analysis of the $20$ item scale and reported four interpretable factors of the CES-D scale as: \textit{ a) depressed affect, b) low positive affect, c) somatic complaints, and d) interpersonal problems}. Based on the loadings of the 20 items on these factors (see Table 11 in \cite{radloff1977ces}), Table \ref{cesd_details} categorizes the questions into four groups. Just like the calculation of the modified CES-D score, we now similarly obtain a sub-scale score for each of these four groups and use them as four outcomes. \cite{hays1998social}, in their study of social correlates of the dimensions of depression, used these four sub-scale scores as the dependent variables. As mentioned in \cite{hays1998social}, and also depicted in Table \ref{cesd_details}, 4 of the 20 items either did not load on any factor or loaded on more than one factor and hence they were excluded from the computation of the sub-scale score.\\
    \\
    \noindent \textbf{\textit{Psychological well-being}}: Developed by the psychologist Carol D. Ryff (\cite{van2008ryff}; adapted from \cite{ryff1989happiness}), the 42-item Psychological well-being (PWB) scale measures six aspects of well-being and happiness: \textit{autonomy, environmental mastery, personal growth, positive relations with others, purpose in life, and self-acceptance}. Each of these six aspects consists of 7 items (thus 42 in total, see Table \ref{pwb_details} and WLS data) and the respondents score how strongly they agree or disagree with each of those items using a 6-point scale (1 = agree strongly; 6 = disagree strongly). Thus, the six aspects will correspond to six different sub-scale score, each varying over the range of 1 ( the respondent has replied only one item under that aspect and the score is 1) to 42 (the respondent has replied all the 7 items under that aspect and the score is 6 for each of them). Table \ref{pwb_details} summarizes the distribution of 42 items over six aspects. For our study, we use these six sub-scale scores along with the the combined score, obtained by summing over all the six sub-scale scores, as the outcomes. For these scores, a higher score implies a better psychological condition.\\

    \noindent \textbf{\textit{Physical health}}: We use SF-12 Physical Component Summary (PCS-12) in the WLS database as one of the outcomes to measure the physical health condition of the woman. In general, SF-12 measure is a multipurpose short-form (SF) generic measure of health status, which is a shorter version of SF-36. Items in SF-12 capture eight concepts (see \cite{ware2019sf}) commonly represented in widely used surveys: \textit{physical functioning, role limitations due to physical health problems, bodily pain, general health, vitality (energy/fatigue), social functioning, role limitations due to emotional problems, and mental health}. The respondents are asked to report the frequencies of each of the items in the past four weeks and based on their answers, the measure Physical Component Summary (PCS-12) is constructed using the physical regression weights summarized in Table 4.1 in \cite{ware2019sf} (see page 22 of \cite{ware2019sf} for more details on the computation of PCS-12). In addition to PCS-12, we also use the number of packs of cigarette the respondents usually smoke per day (or, used to smoke) as another outcome.\\
    \\
    \noindent \textbf{\textit{Alcohol usage}}: : As discussed in \cite{vogelsang2020let}, we consider three measures of potential alcohol abuse: \textit{Possible alcohol dependence, Drinking days per month, and Regular ``at risk" drinker}. What follows next, is the description of the above three measures.
    
    ``Drinking days per month" is the answer to the question “During the last month, on how many days did you drink alcoholic beverages?”. ``Regular at risk drinker" is a binary variable that is coded 1 if the answer to the question “What is the average number of (alcoholic) drinks you had on the days you consumed any alcoholic beverages in the past month?” was $\geq3$ and the respondent drank at least four times over the past month. Finally, ``Possible alcohol dependence" is again a binary variable that is coded as 1, if the respondents admitted to their drinking ever resulting in two or more of the following five consequences: guilt, criticism by others, work-related problems, family problems, or seeking help for their drinking.\\
    \\
    \noindent \textbf{\textit{Economic well-being}}: To measure the economic well-being, we use the Federal percentage of poverty level, which is calculated as dividing the total household income by the poverty guideline and then multiplying that by 100. While the total household income is available in WLS database, poverty guidelines (depending on the number of family members) are available in the ASPE website.

\section{The  design of the data analysis}
\label{testing}
We plan to identify the outcomes that are affected by the treatment, an unintended pregnancy, in both the Catholic subgroup and the non-Catholic subgroup.  Specifically, we aim to discover the outcomes for which the effect of the treatment is negative in both subgroups, as well as the outcomes for which the effect of the treatment is positive in both subgroups. These discoveries are {\em replicable} findings, since the positive (or negative) effect is discovered in both subgroups \footnote{The  direction of  effect  can be measured in various ways. Thus, a positive effect can have several meanings, e.g.,  that  with treatment the outcome: tends to be stochastically larger; or is larger in expectation; or is larger in expectation in at least one of the subscale measures if the hypothesis tested is composed of several subscales hypotheses. The meaning of the positive effect thus depends on the  test statistic used for the inference. (Establishing that there is an effect, without effect direction, does not depend on the  test statistic used for the inference.)  }.
An outcome for which the effect is present only in one subgroup, or for which the effect on that outcome is positive in one subgroup but negative in the other subgroup, is a {\em non-replicable} outcome. Therefore for the purpose of obtaining replicable findings, if such a non-replicable outcome is in our list of discoveries, it is a false positive finding. 
We shall suggest procedures that guarantees that the probability of at least one false positive, i.e.,   the FWER is at most $0.05$.

  Our inference will be based on matched pairs of women, where the matching is made separately for Catholic and non-Catholic women, as described in Section \ref{matching}.  We hereafter follow the notation for causal inference in paired randomized experiments, given in \cite{cross-screening}. For each outcome $k\in\{1, \ldots, 18\}$ we observe its values for $I_{k}^C$ matched pairs of Catholic women, and $I_{k}^{NC}$ matched paired of non-Catholic pairs. We use subscript $k$ because for each outcome we have different numbers of missing values, therefore the number of matched pairs may differ across  outcomes (see Tables \ref{nmpc} and \ref{nmpnc} in the Appendix). 
  
    Consider outcome $k,$ and a matched pair $i\in \{1, \ldots, I_{k}^c\}$ of two Catholic women, $j=1,2.$ The pair is associated with a certain time $t^*,$ such that one of the women had her first unintended pregnancy at time $t^*,$ and is referred to as treated with $Z_{ij}^c=1,$ and the other one had no unintended pregnancies until time $t^*,$ and is referred to as control, with $Z^c_{ij}=0,$ so $Z^c_{i1}+Z^c_{i2}=1.$ 
    The $j$th woman in pair $i$ has two potential values for outcome $k$, $r^c_{T_{ijk}}$ if this woman is treated with $Z_{ij}^c=1$ and $r^c_{C_{ijk}}$ if this woman is not treated, with $Z_{ij}^c=0,$ so the observed value of outcome $k$ for this woman is $R^c_{ijk}=Z_{ij}r^c_{T_{ijk}}+(1-Z_{ij})r^c_{C_{ijk}}.$ The unobserved treatment effect on outcome $k$ for woman $j$ in pair $i$ is $$\delta^c_{ijk}=r^c_{T_{ijk}}-r^c_{C_{ijk}}.$$
  We use similar  definitions addressing outcome $k$ and pair $i'\in\{1, \ldots, I_k^{nc}\}$ of non-Catholic women, replacing superscript $c$ indicating Catholics by superscript  $nc,$ indicating non-Catholics. 

For each outcome,  Fisher's  sharp null hypothesis \citep{fisher1935design} for the Catholics and for the non-Catholics is
  \begin{align} &H_{0k}^c: \delta_{ijk}^c=0 \text{ for } i=1, \ldots, I_k^c, j=1,2, &H_{0k}^{nc}: \delta_{ijk}^{nc}=0 \text{ for } i=1, \ldots, I_k^{nc}, j=1,2. \label{Fisher-null-hyp}\end{align}
  
There are well established methods for testing Fisher's  sharp null hypothesis. 
Let $Z_k^c = (Z_{11k}^c, \ldots, Z^c_{I_K^c2k})$ and $Z_k^{nc} = (Z_{11k}^{nc}, \ldots, Z^{nc}_{I_K^{nc}2k})$ be the vectors of treatment assignments for the Catholic and non-Catholic subgroup, respectively, and  
  $R_k^c = (R_{11k}^c, \ldots, R^c_{I_K^c2k})$ and $R_k^{nc} = (R_{11k}^{nc}, \ldots, R^{nc}_{I_K^{nc}2k})$.  Let $T^c_k = t_k(Z_k^c, R_k^c)$ and $T^{nc}_k = t_k(Z_k^{nc}, R_k^{nc})$ be the test statistics for the Catholic and non-Catholic subgroup, respectively. Different test statistics can be considered for the analysis, among them the popular permutation t-test statistic and Wilcoxon's signed rank test statistic. 
   We use randomization inference to compute the one-sided $p$-values for each subgroup. Assuming Fisher's sharp null hypothesis is true,  the left sided $p$-value towards a negative effect is the probability of receiving at most the value of the observed test statistic, and the right sided $p$-value towards a positive effect is  the probability of receiving at least the value of the observed test statistic. Let $(p_k^c,q_k^c,\tilde{p}_k^c)$ and $(p_k^{nc},q_k^{nc},\tilde{p}_k^{nc})$ denote the triplet left-sided $p$-value, right-sided $p$-value, and minimum between the left-sided and right-sided $p$-value, for outcome $k$  for the Catholic and non-Catholic subgroup, respectively. Different test statistics lead to different $p$-values, but the dependence of the $p$-value on the test statistic, $t_k$, is suppressed for notational convenience. 
  
  We use the sensitivity analysis model described in \cite{cross-screening} to compute the $p$-values after adjustment for the possibility that unobserved biases causes the odds of treatment within a pair to be at most a factor $\Gamma > 1$ ($\Gamma =1 $ corresponds to the randomization inference). We denote by $(p_{\Gamma,k}^c,q_{\Gamma,k}^c,\tilde{p}_{\Gamma,k}^c)$ and $(p_{\Gamma,k}^{nc},q_{\Gamma,k}^{nc},\tilde{p}_{\Gamma,k}^{nc})$ the corresponding sensitivity analysis summaries for the inference.   We shall use $\Gamma= 1.2$ and $\Gamma = 2.0$ for the sensitivity analysis of our replicability analysis findings.  
  
 Our primary aim is to identify the replicated findings.  Thus,
   for each outcome $k\in \{1,\ldots,18\}$, we aim to test  the composite null hypothesis which is the union of Fisher's null hypotheses for the Catholics and the non-Catholics. Rejecting this composite null, we may conclude that there is a treatment effect for both Catholics and non-Catholics. By using right-sided $p$-values, we  conclude that  there is a positive effect for both subgroups. Similarly, by using  left-sided $p$-values, we  conclude that  there is a negative effect for both subgroups. 
   
    We shall show how to provide the replicable findings with a FWER control guarantee, by letting the data guide the choice of which one-sided $p$-value to use, along with other design decisions.   
   Briefly, our suggested approach is non-random cross-screening,  for which the method of \cite{cross-screening} for assessing replicability with FWER control is similar to that of \cite{repl_biometrika}. The idea is that one part of the data  is used to select the hypotheses to be tested based on the second part of the data, and to design the corresponding tests. As noted by \cite{cross-screening}, when the same team of investigators works on both parts of the data, they should decide before seeing the data how one part of the data will be used to design the analysis of the second part. Otherwise, the analysis cannot provide an FWER control guarantee (due to `double dipping', using the same data for both design and analysis). However, if there are two teams of investigators, and each team has access only to one part of data, then each team can  decide how to plan the analysis of the data the other team has access to. This may increase the power of the cross-screening: possibly after seeing their part of the data and  exploratory analyses, the investigators can  suggest a plan for the second team which is better than a plan they would suggest without seeing their part of the data. We will refer to the former cross-screening method as ``Automated cross-screening," and to the latter one, which requires that each team sees only one part of the data, as ``Flexible cross-screening." These two methods are detailed next.       
  \subsection{Automated cross-screening}\label{automated}
    \newlist{steps}{enumerate}{1}
\setlist[steps, 1]{wide=0pt, leftmargin=\parindent, label=Step \arabic*:, font=\bfseries}
      \begin{steps}
      \item [Step 1:] In this step we use the data for the Catholics to select the direction of the alternative regarding each outcome for the non-Catholics, and to select the  hypotheses which will be tested for the non-Catholics.  Finally, we test the selected hypotheses for the non-Catholics. 
          \\\textbf{Selection of directions of alternative for the non-Catholics:} For each outcome $k\in\{1,\ldots,18\},$ compute the Wilcoxon's signed rank test statistic for the Catholic subgroup, and the corresponding right-sided and left-sided $\Gamma$-sensitivity-adjusted $p$-values, $p_{\Gamma,k}^c$ and  $q_{\Gamma, k}^c,$ respectively.
     Select the right-sided alternative for outcome $k$  for the non-Catholics if $p_{\Gamma, k}^c<q_{\Gamma, k}^{c},$ and select the left-sided alternative otherwise.
      Define $D^c_{k}=1$ if right-sided alternative was selected for outcome $k,$ and  $D^c_{k}=-1$ otherwise.
      \\\textbf{Selection of hypotheses for the non-Catholics:}
     For each outcome $k\in\{1, \ldots,18\},$ let $\tilde{p}_{\Gamma, k}^c=\min(p_{\Gamma, k}^c, q_{\Gamma,k}^c)$ be the minimum of the two one-sided $p$-values for the subgroup of the Catholics.  Select the hypotheses to be tested for the non-Catholics based on the $p$-values above as follows: the hypothesis regarding outcome $k$ is selected if $\tilde{p}_{\Gamma,k}^c\leq\alpha/2.$ Let $\mathcal{H}^{nc}\subseteq \{1, \ldots, 18\}$ be the set of indices of selected hypotheses, i.e. $\mathcal{H}^{nc}=\{k: \tilde{p}_{\Gamma, k}^c\leq\alpha/2\}.$ 
     \\ \textbf{Testing the hypotheses for the non-Catholics:}
     For each selected hypothesis with index $k\in \mathcal{H}^{nc},$  compute the one-sided $p$-value $p_{\Gamma, k}^{nc}I(D_k^c=1)+q_{\Gamma, k}^{nc}I(D_k^c=-1),$ where $p_{\Gamma, k}^{nc}$ and $q_{\Gamma, k}^{nc}$ are the right-sided and left-sided  $p$-values for outcome $k$ computed using the data for the non-Catholics, and  $I(\cdot)$ is the indicator. 
     Apply Holm's procedure \citep{holm1979simple} on  
      $\{p_{\Gamma, k}^{nc}I(D_k^c=1)+q_{\Gamma, k}^{nc}I(D_k^c=-1), k\in \mathcal{H}^{nc}\}$ at level $\alpha/2,$ where $\alpha$ is the target FWER level for the replicability analysis. For each rejected hypothesis $H_{0k}^{nc}$ with $k\in\mathcal{H}^{nc},$ define $S^{nc}_k=D_k^{c}.$ For the other hypotheses, which either were not selected for the non-Catholics, or were not rejected, define $S_k^{nc}=0.$ Let $S^{nc}=(S^{nc}_1, \ldots, S^{nc}_{18})$ be the vector indicating the rejection status and the direction of alternative for the hypotheses $(H_{01}^{nc}, \ldots, H_{018}^{nc}).$ 
     \item [Step 2:] In this step we perform Step 1 where the roles of the data for the Catholics and non-Catholics are reversed.
      \item  [Step 3:] For each $k\in\{1, \ldots, 18\}:$ declare outcome $k$ as replicable if and only if both $H_{0k}^c$ and $H_{0k}^{nc}$ were rejected with the same alternative direction, i.e. $S^c_k\times S^{nc}_k=1.$  Claim that the treatment effect on outcome $k$ is positive for both Catholics and non-Catholics if $S^c_k=S^{nc}_k=1,$ and claim that the treatment effect on outcome $k$ is negative for both Catholics and non-Catholics if $S^c_k=S^{nc}_k=-1.$
       \end{steps}
       The above procedure controls the FWER for replicability analysis (see \cite{repl_biometrika}).

  \subsection{Flexible cross-screening for replicability analysis}\label{flexible}
  Having two teams allows for a much more flexible design than the automated cross-screening design. This is because each team can have additional insight from descriptive measures of the subset they are using for designing the analysis in the other subset. In our setting, we aim to design the analysis in the other study for powerful identification of the replicable outcomes, among those listed in Figure \ref{fig:outcomes}.
 Already in the automated cross-screening we gain power by choosing the direction of the hypothesis to test and the reduced set of outcomes tested. Next, we shall detail some aspects of the design that we expect to be enhanced due to the availability of the two teams. 
 
 First, the choice of test statistic to use for each endpoint. In the automated procedure, the test statistic is chosen in advance. However, once we look at the histograms or boxplots of the differences between treated and control groups, we may select another test statistic. 
 In particular,  for the two composite hypotheses that are comprised of the intersection of several of the individual ones, viewing the  distributions of the differences for each outcome that is aggregated into the global score may suggest one aggregation method that is better than others.
 When the hypothesis is comprised of several subscale hypotheses, the team  may chose a test statistic that is sum-based  (e.g., use the sum of the subscale scores, or  use Fisher's combining method) if most subscales seem informative, or a test statistic that is quantile based (e.g., the minimum $p$-value from all subscale tests, or use Simes' combining method) if most subscales seem uninformative. 
 
 The design of the tests may also include additional  information available in the study. For example, the additional waves following survey year 1992-93. In the automated cross-screening, we decided to concentrate on the earliest survey year, where the effect of unintended pregnancy may be strongest. But with the added flexibility of having two teams, there is an opportunity to examine the possibility of designing more complicated tests. 
 It is possible that the team that has access to, e.g., the Catholics subgroup (for the design of the analysis in the non-Catholics subgroup), upon examination of the data for the other survey years (for the Catholics subgroup),  will come to the conclusion that a more powerful design of a test (for the non-Catholics) involves a combination of the outcomes from the different survey years.  

Second, the selection rule for forwarding outcomes for analysis on the other subset may change. In the automated procedure, only $p$-values less than $\alpha/2$ are considered, since  using  Holm it is not possible to establish replicability if    the $p$-value is greater than $\alpha/2$ for that outcome. However, in a non-sparse regime, it may be beneficial to use weights in a weighted Holm procedure and thus  gain power from forwarding hypotheses corresponding to $p$-values that are small enough but larger  than $\alpha/2$. Simulation results in  Appendix \ref{simulation} support this. It is hard to define an automated rule that will decide whether the setting is ``sparse" or ``not sparse" and if ``not sparse," which additional hypotheses to forward for analysis, and to assign to these a reduced weight. But after looking at the data for the subgroup serving for design, a clear picture may emerge. 

Third, related to the second aspect, is the FWER controlling procedure used for the analysis on the selected outcomes. If the team forwards outcomes with $p$-values greater than $\alpha/2$, the multiple testing procedure for FWER control of choice should  incorporate weights in the FWER controlling procedure. 
More generally, designing a closed testing procedure (for the family of individual hypotheses considered for replicability) based on the analysis of the subgroup data the team has access to, is potentially  useful, if the two subgroups have similar signal. 

Finally, since the protocol is written prior to us viewing the data, the flexibility may lead to additional aspects of the design that we haven't anticipated above.   
  
 Let ${\mathcal D}^C$ and ${\mathcal D}^{NC}$ be the observed data in both subgroups.  Flexible cross screening can thus be defined by the following steps:
 \begin{steps}
     \item Team ``NC" uses ${\mathcal D}^C$ to design the inference on ${\mathcal D}^{NC}$. Specifically,  ${\mathcal D}^C$ serves  team ``NC" in order to select the test statistics for each outcome, the direction of the alternative, which hypotheses to forward for inference, and the level $\alpha/2$ FWER controlling procedure to apply on ${\mathcal D}^{NC}$. Similarly, team ``C" uses ${\mathcal D}^{NC}$ to design the inference on ${\mathcal D}^C$.
     \item On ${\mathcal D}^{NC}$, apply the analysis designed by the ``NC" team; on ${\mathcal D}^{C}$, apply the analysis designed by the ``C" team. 
     \item Declare as replicable findings the hypotheses rejected both on ${\mathcal D}^{NC}$ and on ${\mathcal D}^{C}$.
 \end{steps}

 \newcommand{\mP}{\ensuremath{\mathbb P}}
 \newcommand{\mE}{\ensuremath{\mathbb E}}

The design, although random (i.e., not fixed prior to viewing the data), is based only on data that is independent of the data used for inference, since the two subgroups are independent and the two investigative teams act independently. Therefore, the  FWER on false repilcability claims is controlled at level $\alpha$, see Theorem 1 in \cite{repl_biometrika} for the proof. 

   \subsection{Flexible cross-screening for the analysis of global null hypotheses }\label{flexible}

Exploratory analysis is hypothesis generating. Having two independent teams, each examining different subgroups, allows for complete freedom in the exploratory analysis stage for designing the family of hypotheses to be tested in the other subgroup. For replicability analysis, the teams are restricted to a predefined family of hypotheses defined by the list in Figure \ref{fig:outcomes}. But for the analysis of the global null hypotheses, or even more specifically when aiming for inference on each  subgroup separately, this restriction is removed. The added flexibility of having two teams provides the opportunity to consider  more complex null hypotheses, for example:   that  there is no effect modification (moderation) for some of the considered outcomes; that there is no effect in  specific items; that there is no differential (between unintended pregnancy vs. not) decrease from survey year 1992-93 to the subsequent decade (or  no inconsistency across the survey waves). 

Moreover, the survey is more rich than the list of outcomes in Figure \ref{fig:outcomes}. In the exploration step, if unintended pregnancy appears to affect one of these outcomes (e.g., psychological well being), it may encourage the team to generate additional novel hypotheses on related outcomes, following exploration of the correlations between different available outcomes.  For example, if psychological well being was affected, we might be interested in examining whether social participation is affected, as social participation has been found to have connections with psychological well being, see  \cite{sharifian2019differential}.  Another example is as follows. Suppose that the purpose in life scale, which includes questions such as ``To what extent do you agree that you used to set goals for yourself, but that now seems like a waste of time?'', was affected.  Then we might be interested to examine whether a woman's labor force participation was affected by unintended pregnancy, as decreased labor force participation could cause less of a sense of purpose in life, especially if a woman's decreased labor force participation was unintended due to an unintended pregnancy.

In order to have an overall $\alpha = 0.05$ FWER guarantee on all the null hypotheses generated, we shall restrict each team to design an $\alpha/2 = 0.025$ FWER controlling procedure on the novel family of hypotheses generated.  If many hypotheses are of interest, the teams may consider a design that spends the available $\alpha/2$ unequally across hypotheses for better power.  For example, they may consider designing a gatekeeping FWER controlling procedure, where `secondary' hypotheses are tested only if `primary' hypotheses turn out significant. 

For the rejected null hypotheses with well defined parameters, we would like to report effect sizes. For this purpose, we suggest that each subgroup constructs  $\alpha/2$ level  false coverage rate (FCR, \citealt{BY05}) confidence intervals for the selected parameters, thus providing an $\alpha$ level FCR guarantee on all confidence intervals constructed. For example, the team that designs the analysis for the non-Catholic subgroup based on the data from the Catholic subgroup, $\mathcal D^{C}$, will generate $m^{NC} = m(\mathcal D^{C})$ hypotheses, out of which $m^{NC}_{CI}$ may lead to effect sizes of interest. So for each of the $m^{NC}$ hypotheses, the team has to also mention whether a confidence interval will be constructed if this hypothesis is rejected. Let $R^{NC}=R(\mathcal D^{NC})$  be the number of rejected null hypotheses using the  FWER controlling procedure at level $\alpha/2$ on the data for the non-Catholic subgroup $\mathcal D^{NC}$ for the family of $m^{NC}$ hypotheses. Let $R^{NC}_{CI}$ be the number of rejections for which confidence intervals can be constructed, out of the total number of potential confidence intervals (so $R^{NC}_{CI}\leq \min(m^{NC}, R^{NC}$)). By constructing each confidence interval at level $1-R^{NC}_{CI}\alpha/(2m^{NC}_{CI})$, the expected fraction of non-covering confidence intervals (i.e, the FCR) for the subgroup is at most $\alpha/2.$ The theoretical guarantee is exact for independent test statistics, but the procedure  is robust to dependencies encountered  in  practice. 

Two other  stricter procedures to consider, which provide the theoretical coverage guarantee in our setting with dependent test statistics,  are as follows:  the stricter procedure for simultaneous confidence intervals, which constructs each at level $1-\alpha/(2m^{NC}_{CI})$, and guarantees that the probability of non-covering at least one parameter  is at most $\alpha/2$ for any type of dependency; the procedure which constructs confidence intervals at  level $1-R^{NC}_{CI}\alpha/(2{m^{NC}_{CI}\sum_{l=1}^{m_{NC}} \frac 1l}$), and guarantees that the FCR for the subgroup is at most $\alpha/2$ for any type of dependency. The intervals will necessarily be wider, and the stricter error control may not be necessary in our setting, where the purpose is reporting effect sizes with a reasonable confidence guarantee. 

  
 
\bibliographystyle{chicago}
\bibliography{ten_biblio} 
    \appendix

\section{Testing the global null hypotheses}\label{global}
In Section \ref{testing} we described our intended replicability analysis for identifying outcomes which are affected by unintended pregnancy for both Catholics and non-Catholics. In addition to this analysis, we intend to identify the outcomes which are affected by unintended pregnancy for at least one of the subpopulations above. This goal can be addressed by testing the intersection of Fisher's sharp null hypothesis for the Catholics with the same hypothesis for the non-Catholics, with respect to each outcome. Formally, for each outcome $k,$ we are interested in testing the global null $H_{0k}^c\cap H_{0k}^{nc},$ where $H_{0k}^c$ and $H_{0k}^{nc}$ are Fisher's null hypotheses for the Catholics and the non-Catholics respectively, see the definitions in  (\ref{Fisher-null-hyp}). Similarly to our plan regarding replicability analysis (described in Section \ref{testing}), we intend to perform two types of analyses: automated and flexible analyses, where automated analysis  does not require to have two teams working on different parts of the data, while flexible analysis does.
For the automated analysis, we intend to use the following method:       
\setlist[steps, 1]{wide=0pt, leftmargin=\parindent, label=Step \arabic*:, font=\bfseries}
\begin{steps}
   \item For each outcome $k\in\{1, \ldots, 18\},$ use Wilcoxon singed rank test to compute the right-sided p-value and the left-sided p-value for testing whether an unintended pregnancy affects the outcome in the corresponding direction. Let $p_{\Gamma, k}^c, p_{\Gamma, k}^{nc}$ be the $\Gamma$-sensitivity-adjusted right-sided $p$-values for the Catholics and non-Catholics respectively, and let $q_{\Gamma, k}^c, q_{\Gamma, k}^{nc}$ be the left-sided $p$-values for each of the two subpopulations above.
   \item For each outcome $k\in\{1, \ldots, 18\},$ compute the following concordant version of Fisher's global null p-value for testing $H_k^c\cap H_k^{nc},$ following \cite{owen2009karl}, \cite{pearson1934new}:
   $p_{\Gamma, k}^G=2\min (q_{\Gamma, k}, p_{\Gamma, k}),$
           where $$q_{\Gamma, k}=P\left[\chi^2_{(4)}\geq -2(\log(q_{k}^{c})+\log(q_{k}^{nc}))\right],$$  $$p_{\Gamma, k}=P\left[\chi^2_{(4)}\geq -2(\log(p_{k}^{c})+\log(p_{k}^{nc}))\right],$$
   where $\chi^2_{(4)}$ denotes a random variable following a chi-square distribution with 4 degrees of freedom.
    \item Apply Holm's procedure at level $\alpha$ on $\{p_{\Gamma, k}^G, k=1, \ldots, 18\}.$ Let $\mathcal{R}$ be the set of indices of rejected hypotheses.
     \item For each $k\in \mathcal{R},$ reject the global null $H_{0k}^c\cap H_{0k}^{nc},$ i.e. claim that  outcome $k$ is affected by unintended pregnancy for at least one among the two subpopulations.
\end{steps}
   Since the data for the Catholics is independent of the data for the non-Catholics, both $p_{\Gamma,k}$ and $q_{\Gamma, k}$ are valid $p$-values obtained using Fisher's combination method. Following \cite{owen2009karl}, for each outcome $k\in\{1, \ldots, 18\},$  $p_{\Gamma, k}^G$ is a valid $p$-value for testing the global null $H_{0k}^c\cap H_{0k}^{nc},$ therefore Holm's procedure applied on these $p$-values guarantees FWER control with respect to global null discoveries in Step 4. In other words, the method above guarantees that the probability that at least one erroneous claim is made in Step 4 is upper bounded by $\alpha.$ The choice of this method for the automated analysis is based on our simulation results, which show that this method is more powerful than the competitors we considered (see Section \ref{simulation}).
   
   For the flexible analysis, we intend to apply flexible cross-screening described in Section \ref{flexible}, where in the last step, the rejected global null hypotheses are those that are rejected by either one of the teams, rather than those rejected by both teams. This strategy follows the general method of \cite{cross-screening} which has proven FWER control with respect to the global null discoveries.

   \section{Simulation study}\label{simulation}
   We performed a small simulation study in order to compare the power performances of several possible methods addressing replicability and global null analyses. This study provides some insights which helped us to choose methods for the automated analysis and to choose possible design strategies for each of the teams in the flexible analysis.
   The data generation mechanism and the methods considered are given below.
   \paragraph{Data generation and $p$-value computation}
  Motivated by our data structure, this study addresses testing 16 simple null hypotheses for each of two subpopulations, indexed by $j=1,2.$ Hypothesis $H_{kj}$ addresses equality of distributions of outcome $i$ for the treated and for the controls belonging to subpopulation $j,$ for $k=1, \ldots, 16, j=1,2.$  
   Each hypothesis $H_{kj}$ was assigned a truth status for each of the subpopulations, indicated by a vector $(h_{k1}, h_{k2})\in\{(0,0), (0,1), (1,0), (1,1)\},$ where $h_{kj}=0$ if $H_{kj}$ is true and $h_{kj}=1$ otherwise. For each hypothesis $H_{kj}$, we generated a sample of $I$ treatment-minus-control differences $D_{ikj}\sim N(\mu_{kj}/\sqrt{I}, 1),$ $i=1, \ldots, I,$ where $\mu_{kj}=\mu\times h_{kj}$ and $\mu$ is a parameter of the simulation. The left-sided and right-sided $p$-values were obtained based on permutational $t$-test with a sensitivity analysis parameter $\Gamma,$ which was an additional simulation parameter. For ease of notation, we suppress the subscript $\Gamma.$  Let $(p_{k}^{(j)},   q_{k}^{(j)})$ be the right- and left-sided $p$-values for $H_{kj},$ addressing outcome $k$ for subpopulation $j,$ for $k=1, \ldots, 16, j=1,2.$
   \paragraph{Considered methods}
   We addressed separately the goal of replicability analysis (formulated in Secion \ref{testing}), and the goal of global null testing (formulated in Section \ref{global}). All the considered methods have proven FWER control. 
   
   For replicability analysis at level $\alpha,$ we considered a generalized version of automatic cross-screening defined in Section \ref{automated}, which utilizes weights computed based on one part of the data for testing hypotheses based on the other part.
   We refer to this procedure as \textit{weighted cross-screening-replicability}:
\setlist[steps, 1]{wide=0pt, leftmargin=\parindent, label=Step \arabic*:, font=\bfseries}
\begin{steps}
 \item Select the direction of testing for one part based on the other part, as follows. For each $k\in \{1, \ldots, 16\},$ if the test statistic for $H_{k1}$ (namely, the mean of the treatment-minus-control differences $\sum_{i=1}^{I} D_{ik1}/I$) is positive, define $D_k^{(1)}=1$ and select the right-sided alternative for  $H_{k2};$ otherwise, define $D_k^{(1)}=-1,$ and select the left-sided alternative for $H_{k2}.$ Select the direction of the alternative for $H_{k1}$ similarly, based on the sign of the test statistic for $H_{k2},$ and define $D_k^{(2)}=1$ if the test statistic for $H_{k2}$ is positive, and  $D_k^{(2)}=-1$ otherwise. Let $p'_{k1}=p_{k1}I(D_k^{(2)}=1)+q_{k1}I(D_k^{(2)}=-1)$ and $p'_{k2}=p_{k2}I(D_k^{(1)}=1)+q_{k2}I(D_k^{(1)}=-1)$ be the one-sided $p$-values for $H_{k1}$ and $H_{k2},$ respectively, computed for the selected direction of the corresponding alternatives.  
 \item Compute the weights for the hypotheses addressing one population based on the data for the hypotheses addressing the other population as follows. Let $c\in[0,1]$ be a parameter of the simulation. Then the weight for $H_{k1}$ is $w_{k1}(c)=16/(|\{i: p'_{k2}\leq \alpha/2\}|+c|\{i: p'_{k2}>\alpha/2\}|,$ for $k=1, \ldots, 16.$ Similarly, the weight for $H_{k2}$ is $w_{k2}(c)=16/(|\{i: p'_{k1}\leq \alpha/2\}|+c|\{i: p'_{k1}>\alpha/2\}|,$ for $k=1, \ldots, 16.$
 \item Apply a variant of Holm's procedure with weights (referred to as wHolm, see \cite{benjamini1997multiple}) at level $\alpha/2$ on  $\{p'_{k1}, k=1, \ldots, 16\},$ where $H_{k1}$ is associated with weight $w_{k1}(c),$ for $k=1, \ldots, 16.$ Similarly, apply wHolm at level $\alpha/2$  on $\{p'_{k2}, k=1, \ldots, 16\},$ where $H_{k2}$ is associated with weight $w_{k2}(c),$ $k=1, \ldots, 16.$ \item Reject the no replicability null hypothesis $H_{k1}\cup H_{k2}$ if both $H_{k1}$ and $H_{k2}$ were rejected.
\end{steps}
The wHolm procedure is defined as follows. Let $H_1, \ldots, H_m$ be the hypotheses associated with weights $w_1, \ldots, w_m,$ satisfying $\sum_{i=1}^m w_i=m.$ Let $p_{(1)}\leq p_{(2)}\leq \ldots\leq p_{(m)} $ be the ordered sequence of $p$-values for these hypotheses, and let $H_{(i)}, \,w_{(i)}$ be the hypothesis and the weight corresponding to $p_{(i)}.$ Reject $H_{(i)}$ if $p_{(j)}\leq w_{(j)}\alpha/\sum_{k=j}^{m} w_{(k)}$ for all $j\in \{1, \ldots, i\}.$ \cite{benjamini1997multiple} proved that this procedure guarantees FWER control at level $\alpha.$ 
Note that the weighted cross-screening-rep procedure reduces to the automated cross-screening in Section \ref{testing} for $c=0,$ and it reduces to applying Holm's procedure on the entire set of one-sided $p$-values for $c=1.$ 

The weighted cross-screening-replicability procedure was compared to applying Holm's procedure at level $\alpha$ on $\{p_k^{max}, k=1, \ldots, 16\},$ where $$p_k^{max}=2\min\left[\max(p^{(1)}_{k},p^{(2)}_{k}), \max( q^{(1)}_{k},q^{(2)}_{k})\right].$$ This is a  valid $p$-value for $H_{k1}\cup H_{k2}$ (see \cite{benjamini2008screening}, \cite{owen2009karl}). We refer to this competitor as \textit{Holm on maximum $p$-values}.

\begin{figure}[h!]
		\centering
	
			\includegraphics[width=15cm, height=18.5cm]{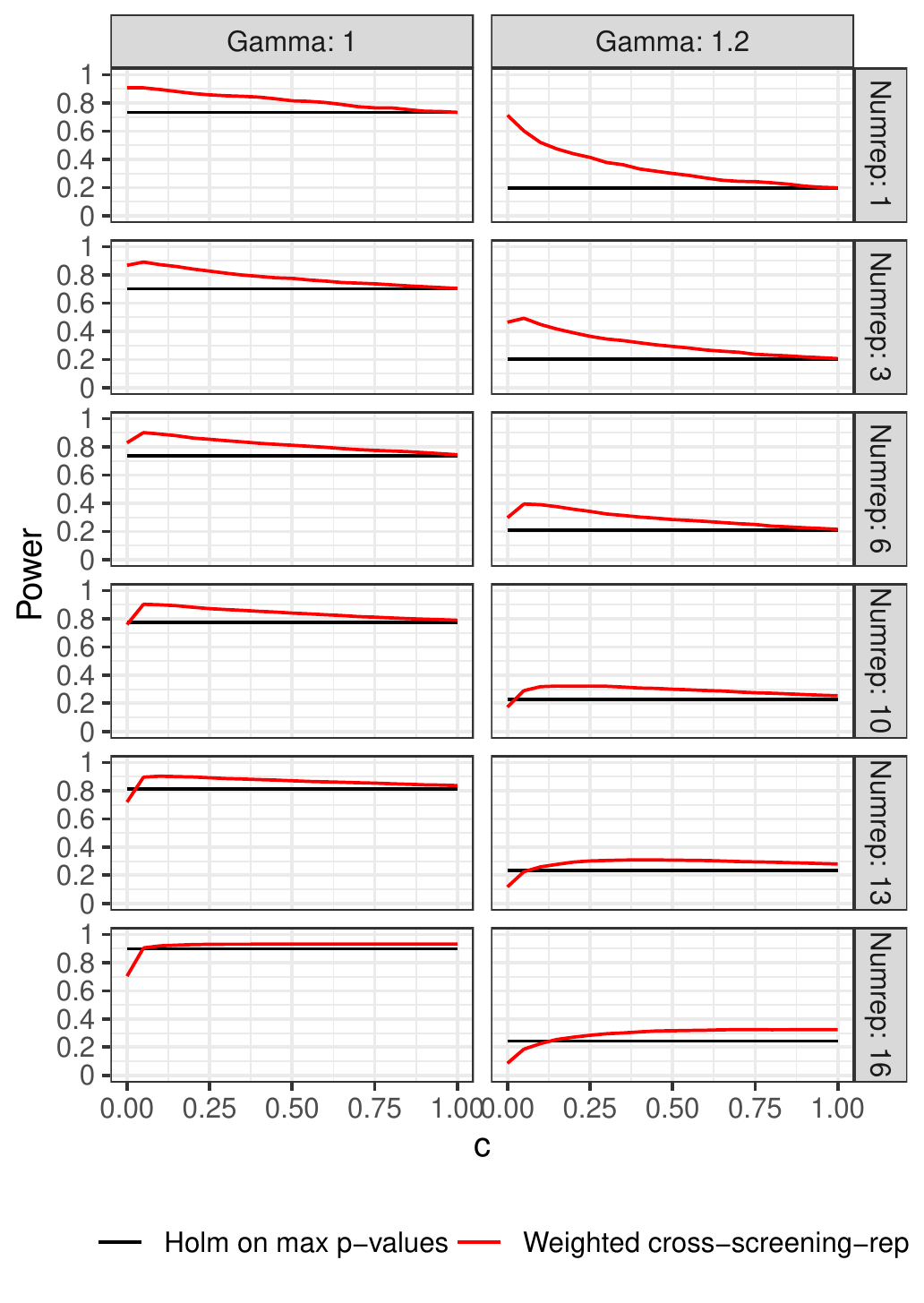}
	\caption{The estimated replicability power for weighted cross-screening-replicability as a function of the parameter $c\in [0,1],$ and for Holm on maximum $p$-values (which does not depend on $c),$ for $\mu=4.$ The rows correspond to different numbers of replicable effects, $K_{11}\in \{1,3,6,10, 13, 16\}.$ The columns correspond to $\Gamma\in \{1, 1.2\}.$}\label{fig:rep}
	\end{figure}

For level-$\alpha$ global null testing, we considered the following methods:
\begin{itemize}
\item \textit{Weighted cross-screening-global:} The method performs Steps 1--3 of the weighted cross-screening-rep method described above, and in Step 4 it rejects the global null hypothesis $H_{k1}\cap H_{k2}$ if at least one among the hypotheses $H_{k1}, H_{k2}$ was rejected in Step 3. In other words, rather then rejecting the hypotheses belonging to the intersection of the two rejection sets obtained in Step 3, it rejects the hypotheses in the union of these two sets.  
    \item \textit{Holm for global nulls:} The method  applies Steps 2-4 given in Section \ref{global} where $p_{k}^c,$ $p_{k}^{nc},$ $q_{k}^c,$ and $q_{k}^{nc}$ are replaced by $p_{k}^{(1)},$ $p_{k}^{(2)},$ $q_{k}^{(1)},$ $q_{k}^{(2)},$ respectively. 
    \item \textit{Holm:} Let $\tilde{p}_k^{(j)}=\min(p_{k}^{(j)}, q_{k}^{(j)}).$ The method applies Holm's procedure at level $\alpha$ on the entire set of two-sided $p$-values, $\{2\tilde{p}_1^{(1)}, \ldots, 2\tilde{p}_{16}^{(1)}, 2\tilde{p}_1^{(2)}, \ldots, 2\tilde{p}_{16}^{(2)}\},$ and rejects $H_{k1}\cap H_{k2}$ if either $H_{k1}$ or $H_{k2}$ was rejected.
    \end{itemize}
\paragraph{Simulation results}
We considered the configurations were all the effects were replicable, i.e. for each $k\in \{1, \ldots, 16\},$ $(h_{k1}, h_{k2})\in \{(0,0), (1, 1)\}.$ The number of replicable effects, $K_{11}=|\{k: (h_{k1}, h_{k2})=(1,1)
\}|$ varied from 1 to 16, and $\Gamma$ varied from 1 to 1.5. Figure \ref{fig:rep} presents the estimated power for weighted cross-screening-replicability as a function of its parameter $c,$ along with the estimated power of its competitor, Holm on maximum $p$-values, for $\Gamma\in\{1, 1.2\}$ and $K_{11}\in\{1, 3,6,10, 13, 16\}.$ For each of the configurations, there exists a range of values of $c$ for which the weighted-cross-screening replicability method is more powerful than its competitor. For the sparse settings, i.e. when the number of replicable effects is small, the weighted cross-screening-replicability method achieves its highest power for $c=0$ or for $c$ which is close to 0, and for these values of $c$ it may be far more powerful than its competitor. For example, for $\Gamma=1.2,$ $K_{11}=1,$ and $c=0,$ the power gain of weighted cross-screening-replicability is close to 0.5. As the number of replicable effects increases, the optimal value of $c$ increases. We observed similar behavior for $\Gamma=1.5.$  These results guided us to choose weighted cross-screening with $c=0$ for the automated replicability analysis. As discussed in Section \ref{flexible}, they can be used for gaining power by flexible cross-screening.

\begin{figure}[h!]
		\centering
	
			\includegraphics[width=15cm, height=18.5cm]{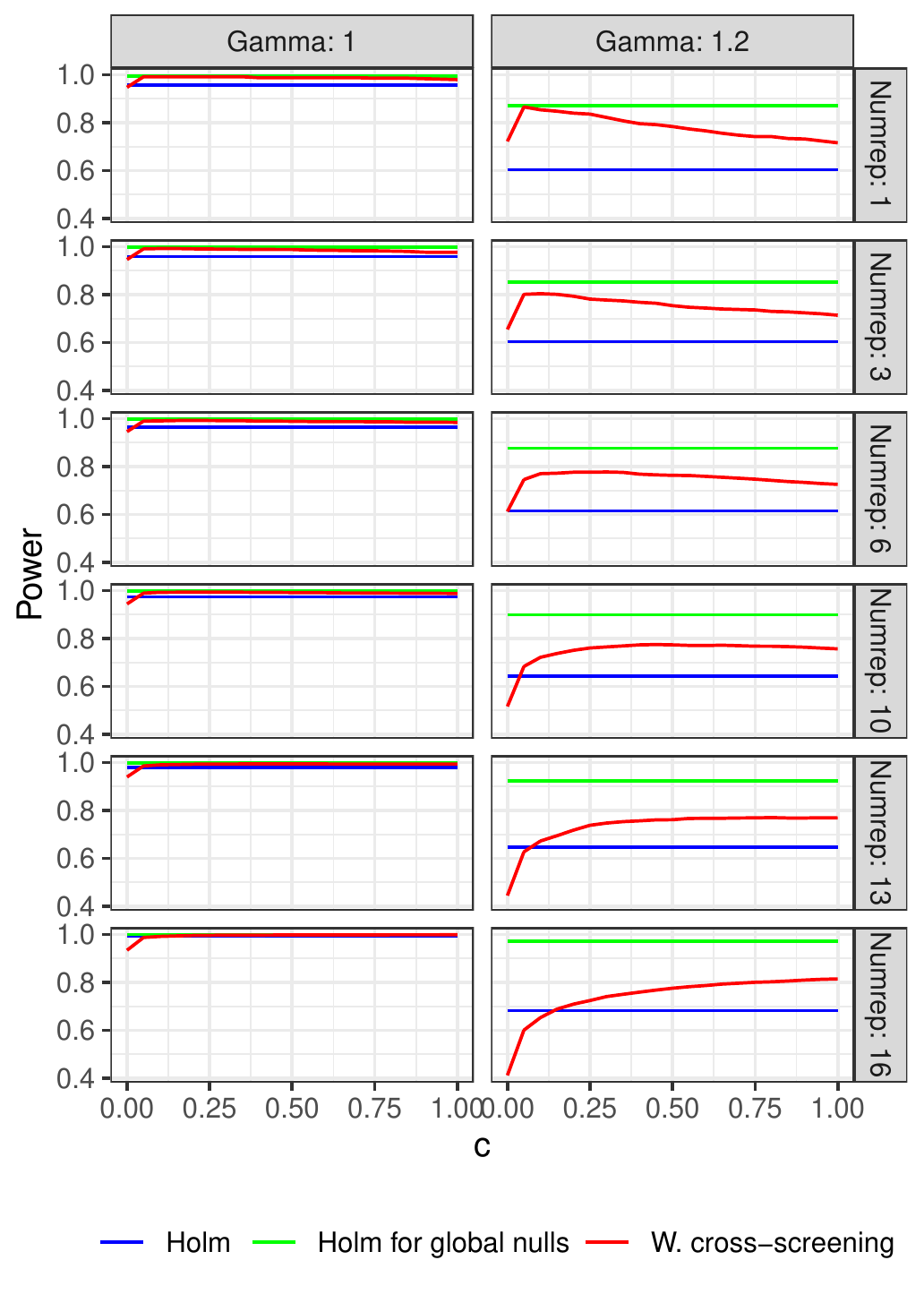}
	\caption{The estimated  power for testing the global nulls by  weighted cross-screening-global as a function of the parameter $c\in [0,1],$ by Holm, and by Holm for global nulls (the latter two methods do not depend on $c),$ for $\mu=3.$ The rows correspond to different numbers of replicable effects, $K_{11}\in \{1,3,6,10, 13, 16\}.$ The columns correspond to $\Gamma\in \{1, 1.2\}.$}\label{fig:global}
	\end{figure}

In addition to replicability analysis methods, we considered  the methods described in Section \ref{global} for testing the global null hypotheses. The estimated power of these methods is given in Figure \ref{fig:global}. It can be seen that the behavior of weighted cross-screening as a function of the parameter $c$ is similar to that in Figure \ref{fig:rep}, however the most powerful method for all the values of $c$ is Holm for global nulls, therefore we choose this method for testing the global nulls in the automated analysis. 
 \clearpage
    \section{Additional Tables}
    \label{add_tebs}
    \begin{table}[ht]
\begin{center}
\scalebox{0.8}{
\begin{tabular}{|c|c|}
    \hline
    \multirow{5}{*}{\textit{Depressed Affect}}&I felt I could not shake the blues\\
    &I felt depressed\\&I felt lonely\\&I had crying spells\\&I felt sad\\
    \hline
    \multirow{3}{*}{\textit{Low Positive Affect}}&I felt as good as other people\\
    &I was happy\\&I enjoyed life\\
    \hline
    \multirow{6}{*}{\textit{Somatic Complaints}}&I did not feel like eating; my appetite was poor\\
    &I had trouble keeping my mind on what I was doing\\&I felt like everything I did was an effort\\&My sleep was restless\\&I talked less than usual\\&I could not get `going'\\
    \hline
    \multirow{2}{*}{\textit{Interpersonal Problems}}&People were unfriendly\\
    &I felt that people disliked me\\
    \hline
    \multirow{4}{*}{\textit{Items not used in sub-scale scores}}&I felt fearful\\
    & I was bothered by things that don't usually bother me\\&I thought my life had been a failure\\&I felt hopeful about the future\\
    \hline
\end{tabular}
}
\end{center}
\caption{\small{Items in CES-D are categorized into four groups (\cite{radloff1977ces}, \cite{hays1998social}): Depressed Affect, Low positive Affect, Somatic Complaints and Interpersonal problems, with each group corresponding to a sub-scale score. Last four items were not used to obtain any of the sub-scale scores (see Section \ref{outcomes}).}}
\label{cesd_details}
\end{table}
    \begin{table}[ht]
\begin{center}
\scalebox{0.75}{
\begin{tabular}{|c|c|}
    \hline
    \multirow{7}{*}{\textit{Autonomy}}&your decisions are not usually influenced\\
    &you have confidence in your opinions \\&you tend to worry about what other people think of you\\&you often change your mind about decisions if your friends or family disagree\\&you are not afraid to voice your opinions\\&being happy with yourself is more important to you \\& it's difficult for you to voice your opinions on controversial matters\\
    \hline
    \multirow{7}{*}{\textit{Environmental Mastery}}&you are good at juggling your time so that you can fit everything in that needs to get done\\
    &you often feel overwhelmed by your responsibilities \\&you are quite good at managing the many responsibilities of your daily life\\&you do not fit very well with the people and community around you\\&you have difficulty arranging your life in a way that is satisfying to you\\&you have been able to create a lifestyle for yourself that is much to your liking \\& you generally do a good job of taking care of your personal finances and affairs\\
    \hline
    \multirow{7}{*}{\textit{Personal Growth}}&you are not interested in activities that will expand your horizons\\
    &you have the sense that you have developed a lot as a person over time \\&when you think about it, you haven't really improved much as a person over the years\\&it is important to have new experiences that challenge how you think about yourself and the world\\&you don't want to try new ways of doing things, i.e. your life is fine the way it is\\&you do not enjoy being in new situations that require you to change your old familiar ways of doing things \\& there is truth to the saying you can't teach an old dog new tricks\\
    \hline
    \multirow{7}{*}{\textit{Positive relation to others}}&you don't have many people who want to listen when you need to talk\\
    & you enjoy personal and mutual conversations with family members and friends\\&you often feel lonely because you have few close friends with whom to share your concerns\\& it seems to you that most other people have more friends than you do\\&people would describe you as a giving person, willing to share your time with others\\&most people see you as loving and affectionate \\& you know you can trust your friends, and they know they can trust you\\
    \hline
    \multirow{7}{*}{\textit{Purpose in life}}& you enjoy making plans for the future and working to make them a reality\\
    & your daily activities often seem trivial and unimportant to you\\& you are an active person in carrying out the plans you set for yourself\\&  you tend to focus on the present, because the future nearly always brings you problems\\&you don't have a good sense of what it is you are trying to accomplish in life\\&you sometimes feel as if you've done all there is to do in life\\& you used to set goals for yourself, but that now seems like a waste of time\\
    \hline
    \multirow{7}{*}{\textit{Self-acceptance}}& you feel like many of the people you know have gotten more out of life than you have\\
    & in general, you feel confident and positive about yourself\\& when you compare yourself to friends and acquaintances, it makes you feel good about who you are\\&  your attitude about yourself is probably not as positive as most people feel about themselves\\&you made some mistakes in the past, but you feel that all in all everything has worked out for the best\\& the past had its ups and downs, but in general, you wouldn't want to change it\\& in many ways you feel disappointed about your achievements in life\\
    \hline
\end{tabular}
}
\end{center}
\caption{\small{Items in Psychological well-being are categorized into six aspects. Each sub-scale score corresponding to each of the aspects is used as secondary outcome and the overall score, that is the sum of six sub-scale score is used as the primary outcome.}}
\label{pwb_details}
\end{table} 

\begin{table}
\centering
\begin{tabular}{|c | c| c| c| c|}
\hline
\textit{Variable} & \textit{Year-1992} & \textit{Year-2003} & \textit{Year-2011} \\
\hline
CESD&229 &182 &117 \\
Income&299&214&161 \\
Alcohol&93&87&71\\
SF-12&Unavailable&153&97\\
Smoking&224&179&121\\
PWB&188&201&145\\
\hline
\end{tabular}
\caption{Number of Catholic matched-pairs with non-missing values for each of the variables.} \label{nmpc}
\end{table}

\begin{table}
\centering
\begin{tabular}{|c | c| c| c| c|}
\hline
\textit{Variable} & \textit{Year-1992} & \textit{Year-2003} & \textit{Year-2011} \\
\hline
CESD&306 &245 &149 \\
Income&368&266&201 \\
Alcohol&80&85&62\\
SF-12&Unavailable&206&132\\
Smoking&296&243&152\\
PWB&227&247&162\\
\hline
\end{tabular}
\caption{Number of Non-Catholic matched-pairs with non-missing values for each of the variables.} \label{nmpnc}
\end{table}
\end{document}